\documentclass[traditabstract]{aa}
\usepackage{graphicx}
\usepackage{natbib}
\usepackage{txfonts}
\usepackage{epsfig}
\usepackage{times}

\begin{document}

\title{Probing AGN triggering mechanisms through the starburstiness of the host galaxies}

  \author{A. Lamastra, N. Menci, F. Fiore, P. Santini, A. Bongiorno, E. Piconcelli}
   \offprints{alessandra.lamastra@oa-roma.inaf.it}
   \institute{INAF - Osservatorio Astronomico di Roma, via di Frascati 33, 00040 Monte Porzio Catone, Italy.}
   \date{Received ; Accepted }
   \abstract{We estimate the fraction of AGNs hosted in starburst galaxies ($\rm {\rm {f_{bursty}}}$) as a function of the AGN luminosity predicted under the assumption that  starburst events and AGN activity are triggered by galaxy interactions during their merging histories. The latter are described through Monte Carlo realizations, and are connected to  star formation and BH accretion using a semi-analytic model of galaxy formation in a cosmological framework. The predicted fraction $\rm {\rm {f_{bursty}}}$  increases steeply with AGN  luminosity  from $ \lesssim $0.2 at $\rm {L_X} \lesssim$ 10$^{44}$erg/s to $ \gtrsim$0.9 at $\rm {L_X} \gtrsim$ 10$^{45}$erg/s over a wide redshift interval from $ z\simeq $0 to $ z\simeq $6. We compare the model predictions with new measurements of $\rm {\rm {f_{bursty}}}$  from  a sample of X-ray selected AGNs in the XMM-COSMOS field at 0.3$< z< $2, and from a sample of QSOs ($\rm {L_{X}} \gtrsim$10$^{45}$erg/s) in the redshift range  2$< z< $6.5. We find preliminary indications that under conservative assumptions half of the QSO hosts are starburst galaxies. This result provide motivation for future systematic studies of the stellar properties of high luminosity AGN hosts in order to constrain AGN triggering mechanisms.

   }
\authorrunning{A. Lamastra, et al.}

 \maketitle

\section{Introduction}

The  tight correlations between the black hole (BH) mass  and the properties of the host galaxy spheroid, including  mass, luminosity, and stellar velocity dispersion 
\citep{Kormendy95,Magorrian98,Ho99,Gebhardt00,Ferrarese00,Marconi03,Haring04,Kormendy09}, imply a tight link between galaxy evolution and BH growth.  Since the growth of BHs is mostly due to accretion of matter during AGN phases \citep{Soltan82}, these findings suggest that the mechanisms responsible for building up the stellar mass in spheroids are also related to the triggering of AGN activity. Theoretical models connecting AGN activity to the merging histories of the host galaxies naturally reproduce the observed BH mass-stellar mass relation in the local \citep[e.g.][]{Peng07,Bower06,Croton06b,Somerville08,Marulli08,Hirschmann10} and higher redshift Universe \citep[e.g.][]{Lamastra10}. 
However, since the mass is an integrated quantity, these correlations could also originate from the hierarchical aggregation of mass \citep[][]{Jahnke11}.

Further hints for the nature of AGN triggering mechanisms could be provided by  the study of  the correlation between the derivative of the BH mass and the stellar mass, namely, the accretion rate and the star formation rate (SFR).
Most of the studies of this correlation have been  done through observations in  the characteristic emission bands of the two processes, namely, the far-infrared emission from cold dust heated by the UV radiation of massive young stars and the hard X-ray emission from the hot corona of the AGN.  Observational studies based on those indicators  revealed a complex situation. In the local Universe \cite{Netzer09} found a strong correlation between the luminosity at  60 $\mu$m  and the AGN luminosity for optically-selected AGNs  over more than five orders of magnitude in luminosity. According to \cite{Rosario12} the luminosity at  60 $\mu$m  is correlated to the AGN luminosity for  AGNs with X-ray luminosities  $\rm {L_{X}}\gtrsim10^{43}$ erg/s at $z<1$, while no correlation is found for AGNs of the same luminosity at higher redshift and for lower luminosity AGNs \citep[see also][]{Lutz10,Shao10}. \cite{Mullaney12a} found no evidence of any correlation between  the X-ray and infrared luminosities of AGN with  $\rm{L_{X}}=10^{42}-10^{44}$ erg/s up to $z=3$. However the correlation arises at $z=$1-2 when stacked X-ray emission of undetected sources is taken into account \citep{Mullaney12b}. Finally, \cite{Rovilos12} found evidence for a correlation between $\rm {L_{X}}$ and the SFR per unit stellar mass of the host galaxy (SSFR=SFR/M$_*$) for  AGNs with $\rm {L_{X}}>10^{43}$ erg/s at $z>1$,  and they did not find evidence for such correlation for lower luminosity systems or those at lower redshifts.

Although this puzzling situation might be related to observational bias and/or AGN variability \citep{Hickox13,Chen13}, a simple explanation is that the AGN activity is linked only to a particular type of star formation \citep{Neistein13}. 
Indeed, there is a growing  observational evidence for two modes of star formation: a quiescent mode that takes place in most star forming galaxies with gas conversion  time scales of $\simeq$1 Gyr, and the less common starburst mode  acting on much shorter time scales of $ \simeq $ 10$^7$ yr \citep[e.g.][]{Rodighiero11,Lamastra13}.  There is a strong  theoretical argument indicating that the latter mode is the one related to the AGN activity, at least for the most luminous AGNs.  In fact,  to achieve high bolometric luminosities of $\rm {L_{bol}}$=$\eta \Delta$ $\rm{M_{gas}}$c$^{2}$/$\Delta$t=10$^{46}$erg/s,  typical of QSOs, in host galaxies with gas content of 5$\times$10$^{8}$ $\rm {M_{\odot}}$, typical AGN activity times $\Delta$t$\leq$a few times 10$^7$yrs are required even for a large destabilized gas mass $\Delta$$\rm{M_{gas}} \simeq $$\rm{M_{gas}}$/5 (where $\eta$=0.1 is commonly assumed for the radiative efficiency). This implies that fuelling a QSO requires the loss of a sizeable fraction of the  disk angular momentum. This must happen on a timescale that is comparable to or shorter than the dynamical time of the galaxy.  At present, galaxy merging seems to be the best (if not the only) mechanism with the above properties.  Indeed, high-resolution numerical simulations have shown the effectiveness of galaxy  major merging in funnelling  a large amount of  gas onto the nuclear regions in  a short time scale, and  the resulting high gas density in the central region of the galaxy  at the same time triggers starburst events \citep{Hernquist89, Barnes91, Barnes96, Mihos94, Mihos96,DiMatteo05,Cox08}.  The connection between AGNs, starburst galaxies and galaxy mergers has several observational confirmations.  Major mergers are associated with enhancements in star formation in local ultra luminous infrared galaxies \citep[ULIRGs,][]{Sanders96, Elbaz07}, and at least some sub-millimetre
galaxies \citep[SMGs,][]{Tacconi08, Daddi07, Daddi09, Engel10,Fu13}. The fraction of galaxies hosting AGN activity is correlated to the IR luminosity \citep{Kim98,Veilleux99,Tran01,Alexander08}. Support for this scenario also comes from the signature of recent mergers in QSO hosts \citep[see, e.g.,][] {Bennert08,Treister12}. 
Based on the above evidence,  semi-analytic models (SAMs) of BH and galaxy evolution assume galaxy major mergers as triggers for QSO activity \citep[e.g.][]{Kauffman00,Menci03,Menci06,Croton06,Bower06,Hopkins06b,Monaco07,Marulli08,Somerville08}.
For less luminous AGNs (Seyfert-like galaxies $\rm {L_{bol}} \lesssim$10$^{45}$ erg/s), different fuelling mechanisms have been proposed in the literature. These include minor mergers, disk instabilities, the stochastic accretion of cold molecular clouds near the BH, and Bondi-Hoyle spherical accretion of hot gas from the diffuse atmosphere in the central bulge \citep[e.g.][]{Fanidakis12,Hirschman12,Bournaud11,Bournaud12}. These processes are often referred to as ``secular processes'' and their connection (if any) with  the star formation of the host galaxies is less clear.
Thus,  a  stronger correlation between SFR and AGN luminosity is expected for more luminous AGNs than for  less luminous sources.

A  direct approach to test the AGN-starburst-merger scenario is to  identify a diagnostic that isolates  the star formation  directly related to the AGN luminosity and study its dependence on AGN luminosity.  
An effective tool for distinguish  between quiescently star forming galaxies and starburst galaxies  is provided by
the scaling relation connecting the SFR with the total stellar mass. It has  been shown that the former lie along a ``main sequence'' characterized
by a typical redshift-dependent value of the SSFR \citep{Brinchmann04, Noeske07, Elbaz07, Daddi07, Daddi09, Santini09, Salim07, Stark09, Gonzalez11, Rodighiero11}, while the less numerous starburst population has higher SSFRs \citep[e.g.][]{Rodighiero11}.
 Thus the comparison between  the observed  distribution of AGN hosts in the SFR-M$_*$ plane and the observed fraction of starbursting hosts as a function of the AGN luminosity with those predicted by  merger-driven models for starbusrts and AGNs represent a test case for their basic assumption about the fuelling mechanism. In this paper we carry out these comparisons using  a SAM of galaxy formation that includes a physical description of starburst and BH growth triggered by galaxy interaction during their merging histories \citep{Menci08} and  a sample of X-ray selected AGNs in the XMM-COSMOS field  in the redshift range 0.3$< z< $2, and  a sample of QSOs ($\rm {L_{bol}} >$10$^{46}$ erg/s) at  2$< z< $6.5. Our SAM is ideally suited to this goal as it has been tested against the separation of  starburst and quiescently star forming galaxies in the  SFR-M$_*$ diagram as well as against several observational properties of the galaxy and AGN populations \citep{Menci04, Menci05,Menci06,Menci08, Lamastra10,Lamastra13}.

The paper is organized as follow. A description of the SAM  is given in Section \ref{model}; in Section \ref{sample}  we describe the properties of the AGN samples; in Section \ref{results} we derive the predicted  
relation between the SFR and the AGN luminosity and the host galaxy stellar mass and the fraction of starbursting hosts as a function of AGN luminosity;  conclusions follow in  Section \ref{conclusions}.

\section{The model}\label{model}

We adopt the SAM described in details in \cite{Menci04, Menci05, Menci06, Menci08},  which connects, within a cosmological framework,
the baryonic processes (gas cooling, star formation, BH accretion, supernova and AGN  feedback) to the merging histories of the dark matter (DM) haloes. 
The latter, including the gradual inclusion of sub-haloes and their subsequent coalescence due to dynamical friction or binary aggregation, is computed adopting a Monte Carlo technique. 
The properties of the gas and stars contained in the DM haloes are computed following the standard recipes commonly adopted in SAMs. 
Starting from an initial amount $m\Omega_b/\Omega$ of gas at virial temperature of the galactic haloes, we compute the mass $m_c$ of cold baryons which are able to radiatively cool.
The cooled gas mass $m_c$ settles into a rotationally supported disk with radius $r_d$
(typically ranging from $1$ to $5 $ kpc), rotation velocity $v_d$,
and dynamical time $t_d=r_d/v_d$, all computed after \cite{Mo98}. 
Stars form with a rate 
\begin{equation}
 {\rm SFR_{q}} = m_c/\tau_q,
 \label{sfr_q}
\end{equation}
where $\tau_q=qt_d$, and  $q$ is a model free parameter that is chosen to match the \cite{Kennicutt98}  relation. In the following, we will refer to this mode of star formation as quiescent.\\
At each time step, the mass $\Delta m_h$ returned from  the cold gas content of the disk to the  hot gas phase owing to the energy released by SNae following star formation is estimated from canonical energy balance arguments as $\Delta m_h=E_{SN}\epsilon_0 \eta_0 \Delta m_* /v_c^2$, where $ \Delta m_*  $  is the stellar mass formed in the time step, $\eta_0$ is the number of SNe per unit solar mass (for a Salpeter initial mass function $\eta_0 = 6.5 \times 10^{-3} {\rm M_{\odot}}^{-1}$), $E_{SN}=10^{51} erg/s$ is the energy of ejecta of each SN,  $v_c$  the circular velocity of the galactic halo, and $\epsilon_0=0.01$ is a tunable efficiency for the  coupling  of the emitted energy with  the cold interstellar medium.  

\subsection{Starburst events and BH accretion triggered by galaxy interactions}

For a galactic halo with given circular velocity $v_c$ inside a host halo with circular velocity $V$, the interactions occur at a rate
\begin{equation}\label{int}
\tau_r^{-1}=n_T(V)\,\Sigma (r_t,v_c,V)\,V_{rel} (V),
\end{equation}
where $n_T$ is the number density of galaxies in the host halo,
$V_{rel}$ the relative velocity between galaxies, and $\Sigma \simeq \pi\langle r_t^2+r_t^{'2}\rangle$  the cross section for
such encounters, which is  given by  \cite{Saslaw85}  in terms of the tidal radius
$r_t$ associated to a galaxy with given circular velocity $v_c$ \citep[see][]{Menci04}. \\
The fraction $f$ of cold gas destabilized by the interactions has been worked out by  \cite{Cavaliere00}  in terms of the  variation $\Delta j$ of the specific angular momentum $j\approx
Gm/v_d$ of the gas to read  \citep{Menci04}:
\begin{equation}
f\approx \frac{1}{2}\,
\Big|{\Delta j\over j}\Big|=
\frac{1}{2}\Big\langle {m'\over m}\,{r_d\over b}\,{v_d\over V_{rel}}\Big\rangle\, ,
\label{f}
\end{equation}
where $b$ is the impact parameter, evaluated as the greater of the radius $r_d$ and the average distance of the galaxies in the halo, $m'$ is the mass of the  partner galaxy in the
interaction, and the average runs over the probability of finding such a galaxy
in the same halo where the galaxy with mass $m$ is located.  The pre-factor accounts for the probability 1/2 of inflow rather than outflow related to the sign of $\Delta j$.
AGN and starburst events are triggered by all  galaxy interactions including  major mergers  ($ m \simeq m^{'} $),  minor
mergers ($ m \ll m^{'}$), and by  fly-by events.
We assume that in each interactions 1/4 of the destabilized  fraction $f$ feeds the central BH, while the remaining fraction feeds the circumnuclear starbursts  \citep[see][]{Sanders96}.
Thus, the star formation rate in the nuclear region due to interaction-driven burst is given by
 \begin{equation}
{\rm SFR_{b}}=(3/4) f m_c/\tau_{b} .
\label{sfr_b}
\end{equation}
Here the time scale $\tau_{b}$  is assumed to be the crossing time  for the destabilized cold gas component ($t_d$).
This adds to the quiescent star formation rate given in eq. (\ref{sfr_q}). \\
The accreted mass $\Delta m_{acc}=(1/4) f m_c$ into the BH powers
the AGN emission with bolometric luminosity
\begin{equation}\label{Lagn}
 {\rm L_{bol}}={\eta\,c^2\Delta m_{acc}\over \tau_{AGN}} ~.
\label{LAGN}
\end{equation}
The duration of an accretion episode, i.e.,
the timescale for the QSO or AGN to shine, is $\tau_{AGN}=t_d$.
We adopt an energy-conversion efficiency $\eta= 0.1$ (see Yu \&
Tremaine 2002), and derive the X-ray luminosities in the 2-10
keV band (${\rm L_X}$) from the bolometric correction given in \cite{Marconi04}

\subsection{AGN feedback and column density of absorbing gas}\label{feedback}

Our SAM  includes a detailed treatment of AGN
feedback which is directly related to the impulsive, luminous AGN phase (Menci et al. 2006, 2008). This is based on expanding blast wave as a mechanism to propagate outwards the AGN energy injected into the interstellar medium at the center of galaxies. 
The injected energy  is taken to be proportional to the energy radiated by the AGN, 
E = $\epsilon_{AGN}\eta c^2 \Delta m_{acc}$, where  $\epsilon_{AGN}$=5 10$^{-2}$ is the value of the energy feedback efficiency for coupling with the surrounding gas.
All the shock properties depends on this quantity.
The AGN emission is absorbed by the unperturbed amount of gas in the galaxy disk outside the shock.
To calculate the neutral hydrogen column densities  ($\rm{N_{H}}$)  of the unshocked absorbing gas we extract a random line-of-sight angle $\theta$, which defines the disk inclination to the observer. At time $t$ within the interval $\tau_{AGN}$ corresponding to the active AGN, we compute $\rm {N_H}$ corresponding  to the gas outside the shock position along the selected line of sight as
\begin{equation}\label{NH}
\rm {\rm {N_{H}}}=\int_{R_s(t)}^{h/\sin \theta}\rho(r)dl.
 \end{equation}
where $\rm{R_{s}(t)}$ is the shock position after a time $t$ from an AGN outburst, h=r$_d$/15   is the disk thickness \citep[see][]{Narayan02}, and $\rho$ is  the density distribution of the unperturbed gas for which we assumed the form $\rho=\rho_0exp(-r/r_d)$ (where $r$ is the distance from the center of the galaxy) with a cut-off in the direction perpendicular to the disk at $r$=h. The density distribution is normalized so as to recover the total gas mass $m_c$ when integrated over the disk volume \citep[see][]{Menci08}.\\
 
 The model has been tested against several observational properties of the galaxy and AGN  populations such as  the evolution of the galaxy and AGN luminosity functions in different bands,  the local M$_{BH}$-M$_*$ relation, the galaxy bimodal colour distribution, the Tully-Fisher relation, the fraction of obscured AGN as a function of luminosity and redshift, and the relative contribution of starburst and main sequence galaxies to the cosmic star formation rate density  \citep{Menci04, Menci05,Menci06,Menci08, Lamastra10,Lamastra13}.

\section{Data set}\label{sample}
Since we aim at separating the burst mode of star formation from the quiescent one through the value of the SSFR we need AGN samples with measured values of SFR and stellar mass. 

The unprecedented wide and deep  multi-wavelength coverage of the COSMOS field makes it possible to derive the total stellar mass, as well 
as the other stellar parameters,  in statistically representative samples of galaxies through the spectral energy distribution (SED) fitting technique.
In our analysis we use a sample of X-ray selected AGNs from the  XMM-$Newton$ survey of the COSMOS field in the redshift range 0.3$<z<$2 \citep{Scoville07}.   
The XMM-COSMOS catalogue has been presented by \cite{Cappelluti09}, while the optical identifications and multi-wavelength properties have been discussed by  \cite{Brusa10}.
The  X-ray luminosities in the (2-10) keV band  ($\rm {L_X}$) are derived by \cite{Mainieri07,Mainieri11}. 
Whenever possible, the de-absorbed  X-ray luminosities are determined with a proper spectral analysis.   
Otherwise, the absorbing column density are derived from the hardness ratio assuming a given photon index.\\ 
The stellar masses of the XMM-COSMOS AGN host galaxies are derived by \cite{Santini12} and \cite{Bongiorno12} by fitting  the observed SEDs with a two component model based on a combination of AGN and host galaxy templates. \cite{Bongiorno12} used the SED fitting technique in the optical/IR  bands also to derive the SFRs.  This procedure relies on measurement of the  UV light emission from young stars corrected for dust extinction. 
As discussed by the authors, these SFRs are reliable only for obscured AGNs where  the UV range is clean of AGN contamination.\\
Based on the assumption that the AGN contamination in far-infrared (FIR) emission is not dominant over the emission of cold dust heated by young stars, 
we derive the SFRs  of  XMM-COSMOS AGNs  from the dust thermal emission at these wavelenghts. We use the 
160 $\mu$m data  collected by the PACS instrument \citep{Poglitsch10} on board the $Herschel$
Space Observatory \citep{Pilbratt10}, as part of the PACS
Evolutionary Probe \citep{Lutz11} survey.  The  fraction of XMM-COSMOS AGNs in the redshift interval 0.3$<z<$2 detected at 160 $\mu$m is 20\%.
 A number of previous studies support this assumption. Indeed, many authors \citep[e.g][]{Schweitzer06,Netzer07,Lutz08} find a strong correlation between FIR luminosity
and SFR tracers, such as PAH emission features, both in local and high redshift bright AGNs. However, the 160 $\mu$m band corresponds to (53-123) $\mu$m rest-frame wavelength band  in the redshift interval considered in this work, thus at high redshift we are sensitive to warmer dust emission. Nevertheless, this band is not strongly affected by the AGN emission as suggested by the study of
\cite{Rosario12} \citep[see also][]{Santini12}. In  the same redshift interval, they compared the FIR 100 $\mu$m to 160 $\mu$m colour of the XMM-COSMOS AGNs   with that of inactive mass-matched galaxies  finding no significant difference between the two samples.  
Since the typical AGN luminosity of flux-limited samples like those from XMM-COSMOS increases towards higher redshifts, while  the  160 $\mu$m
band traces increasingly shorter rest-frame wavelengths, we cannot, however, exclude a minor contribution from  the AGN emission in the most distant sources used in this work.\\
To estimate the SFR the flux at 160 $\mu$m is fitted with the \cite{Dale02} template library to derive the  total IR luminosity  (L$_{IR}$) integrated between 8 $\mu$m and 1000 $\mu$m. The latter is converted into SFR using the relation SFR[$\rm {M_{\odot}}$/yr]=1.7$\times$10$^{-10}$ $\rm{L_{IR}}$ [$\rm {L_{\odot}}$] \citep{Kennicutt98}.


All the SFRs and stellar masses used in this work are computed using a Salpeter initial mass function (IMF).\\

In order to extend the analysis at higher redshifts and AGN luminosities we collect  from the literature  28 QSOs  with $\rm {L_{bol}} \geq $10$^{46}$erg/s \cite[corresponding to $\rm  {L_{X}}\gtrsim$10$^{44.5}$ erg/s  applying the bolometric correction of][]{Marconi04} at 2$<z<$6.5 \citep{Polletta08,Polletta11,Lacy11,Wang10,Wang13,Solomon05,Coppin08,Shields06,Maiolino07,Gallerani12b}.
In this sample, only three QSOs  have estimates of the host galaxy stellar mass and SFR from SED fitting. 
For the other 21 sources we infer the stellar masses from observations of molecular carbon monoxide (CO) emission lines. 
Indeed, these observations provide valuable constraints on the gas content and dynamical state of these systems. 
Under the assumption that the gas is driven by gravity and is approximately virialized, 
the  dynamical masses of the host galaxy can be estimated as $\rm{M_{dyn}}=R v^2/G \sin^2 i$, where $R$ is the disk radius,  $\rm{v}$ is the circular velocity at the outer disk radius which is measured from the CO line width, and $i$ the inclination angle of the gaseous disk.  The main uncertainty of the dynamical mass is due to the unknown inclination angle i.  
We derive the mean value of the dynamical mass assuming randomly oriented disks with respect to the sky plane  as:
\begin{equation}
<\rm{M_{dyn}}>=\frac{\int_{i_{min}}^{i_{max}}\frac{R v^2}{G \sin^2 i} \sin i di}{\int_{i_{min}}^{i_{max}} \sin i di}
\end{equation}
where $\rm{i_{min}}$ is the minimum disk inclination angle obtained by setting M$_*$=$\rm{M_{dyn}}$-$\rm{M_{gas}}$ $<$10$^{13}$ $\rm {M_{\odot}}$, and $\rm{i_{max}}$ is the maximum inclination angle derived by setting
$\rm{M_{dyn}}$ $>$ $\rm{M_{gas}}$.
We estimate  the molecular gas (H$_2$) mass from the CO line luminosity assuming  a CO intensity-to-gas mass conversion factor of $\alpha_{CO}$=$\rm{M_{gas,mol}}$ /$\rm{L_{CO(1-0)}}$=0.8 $\rm {M_{\odot}}$ (K km s${-1}$ pc$^2$)$^{-1}$, as it  is commonly assumed for ULIRGs and QSO host galaxies \citep[e.g.][]{Solomon05,Wang10}. 
Then we infer the stellar masses as M$_*$=$\rm{M_{dyn}}$-$\rm{M_{gas,mol}}$. We here assume negligible atomic gas (HI) contribution to the total gas mass of high-$z$ QSO hosts. Observational evidences \citep[e.g.][]{Daddi10b,Tacconi10,Geach11} and theoretical arguments \citep{Blitz06,Obreschkow09} indicate that the H$_2$/HI fraction increases with redshift, making us confident of the assumption adopted.
By comparing the relation between $\rm{L_{CO}}$ and the luminosity at (42.5-122.5) $\mu$m rest-frame wavelength band ($\rm{L_{FIR}}$) of the CO-detected QSOs  with that of galaxies without a luminous AGN, \cite{Riechers11}  found that $\rm{L_{FIR}}$   in these QSO is dominated by dust-reprocessed emission from young stars in the host galaxy, rather than the AGN. Following \cite{Riechers11}, we derive their SFRs from  $\rm{L_{FIR}}$ under the conservative assumptions that $\rm{L_{IR}}\simeq\rm{L_{FIR}}$, and that 10\% of $\rm{L_{FIR}}$ is actually powered by the AGN and not the starburst. 

We use the same relation to derive the SFR  also for the remaining 4 sources in this sample.  
To estimate their stellar masses we use the angle-corrected dynamical masses within the singly ionized carbon ([C II]) emitting region from \cite{Wang13}, and the gas masses derived from CO observations \citep{Wang10,Wang13} as described above.

We also include in this sample HS1700+6416 ($\rm {L_{X}}=$4$\times$10$^{45}$ erg/s, \citealt{Lanzuisi12}) for which we measure  SFR of 2454 $\rm {M_{\odot}}$/yr and stellar mass of 3$\times$10$^{11}$ $\rm {M_{\odot}}$ from SED fitting (Bongiorno et al. in prep.).

\section{Results}\label{results}

\subsection{The SFR-$\rm {L_X}$ relation }

\begin{figure}[h!]
\begin{center}
\includegraphics[width=7 cm]{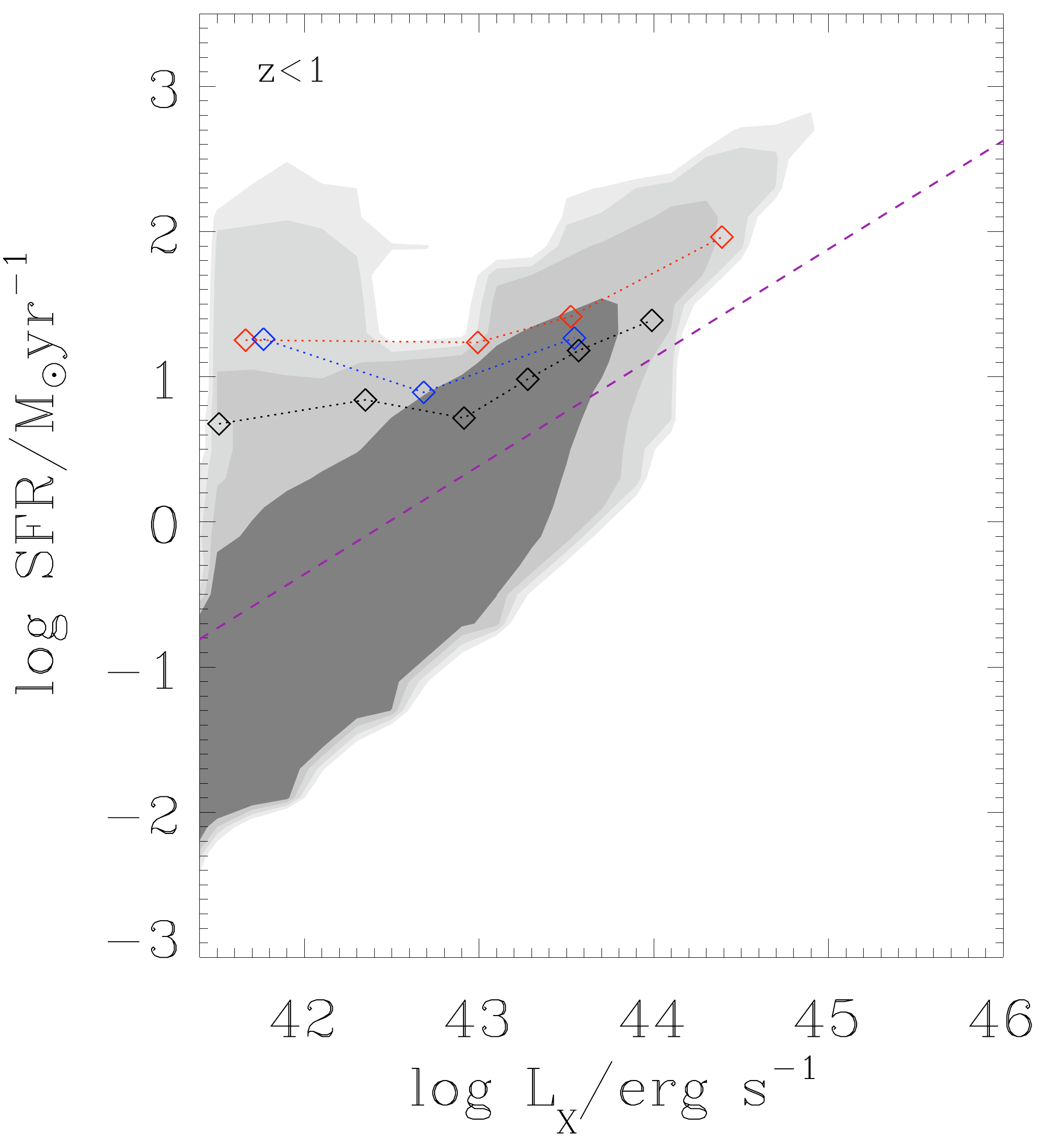}
\caption{SFR  versus $\rm {L_X}$ at $z<$1. The four filled contours correspond to equally spaced values of the density (per Mpc$^3$) of model AGNs  in logarithmic scale: from 10$^{-7}$  for the lightest filled region to 10$^{-4}$  for the darkest. The dashed line is the relation obtained by \cite{Netzer09}. Observational results from \cite{Rosario12} at $z<$0.3, 0.2$<z<$0.5, 0.5$<z<$0.8, are shown in black, blue and red dotted lines, respectively.}
\label{sfr_Lx}
\end{center} 
\end{figure}

We start by showing in figure \ref{sfr_Lx}  the SFR-$\rm {L_X}$ relation of model AGNs at $z<$1. Given the modest redshift evolution of the predicted SFR-$\rm {L_X}$ relation in this redshift range, we represent the whole redshift interval on the same SFR-$\rm {L_X}$ plane  to provide an overview of the trend of SFR as a function of $\rm {L_X}$.
The plotted SFR is the total SFR of the host galaxy which is given by  the sum of the quiescent and burst component of star formation (eq. \ref{sfr_q} and \ref {sfr_b}). Since the accretion rate onto the BHs is correlated only to the latter mode of star formation we find a strong correlation between  SFR and $\rm {L_X}$ for  luminous AGNs, and a more scattered relation for the less luminous sources  owing to the larger contribution of the quiescent component of star formation to the total SFR of the galaxy in these objects.
In figure \ref{sfr_Lx} we also show the observed  relations for local optically-selected AGNs \citep{Netzer09} and for X-ray selected AGNs  at $z<$0.8 \citep{Rosario12}. The data points from  \cite{Rosario12} do not represent individual objects but mean trends that come from  combining fluxes from detections and stacks of undectected  sources in the $Herschel$-PACS bands in bins of  redshift and X-ray luminosity. 
Following \cite{Santini12}, we convert the rest frame luminosity at a wavelength of 60 $\mu$m (as presented in their works) into $\rm{L_{IR}}$  by
linearly fitting the values of $\nu\rm{L_{\nu}(60\mu m)}$ and $\rm{L_{IR}}$  predicted by the \cite{Dale02} templates ($\log\nu\rm{L_{\nu}(60\mu m)}$=1.07$\times\log\rm{L_{IR}}$-1.18, we find a very similar relation using the \citealt{Chary01} template library). We then compute the SFR by using the relation given in Section \ref{sample}.
A detailed comparison with the observational results is beyond the scope of this paper since this requires an accurate estimate of the different selection criteria of the AGN samples. 
Here we note that pinning down the AGN triggering mechanism  from the  SFR-$\rm {L_X}$ relation is a critical issue. On the observational side, the observed relation is still uncertain due to observational bias and instrumental limitation in sensitivity. On the theoretical side, even a model based on galaxy interactions as triggers of AGN and starburst activities, predicts a  large scatter ($\simeq$3 orders of magnitude) at low AGN luminosities  owing to the pollution of the global star formation by the quiescent mode.

\subsection{The starburstiness-$\rm {L_X}$ relation }\label{fbursty}
\begin{figure*}[]
\begin{center}
\includegraphics[width=16 cm]{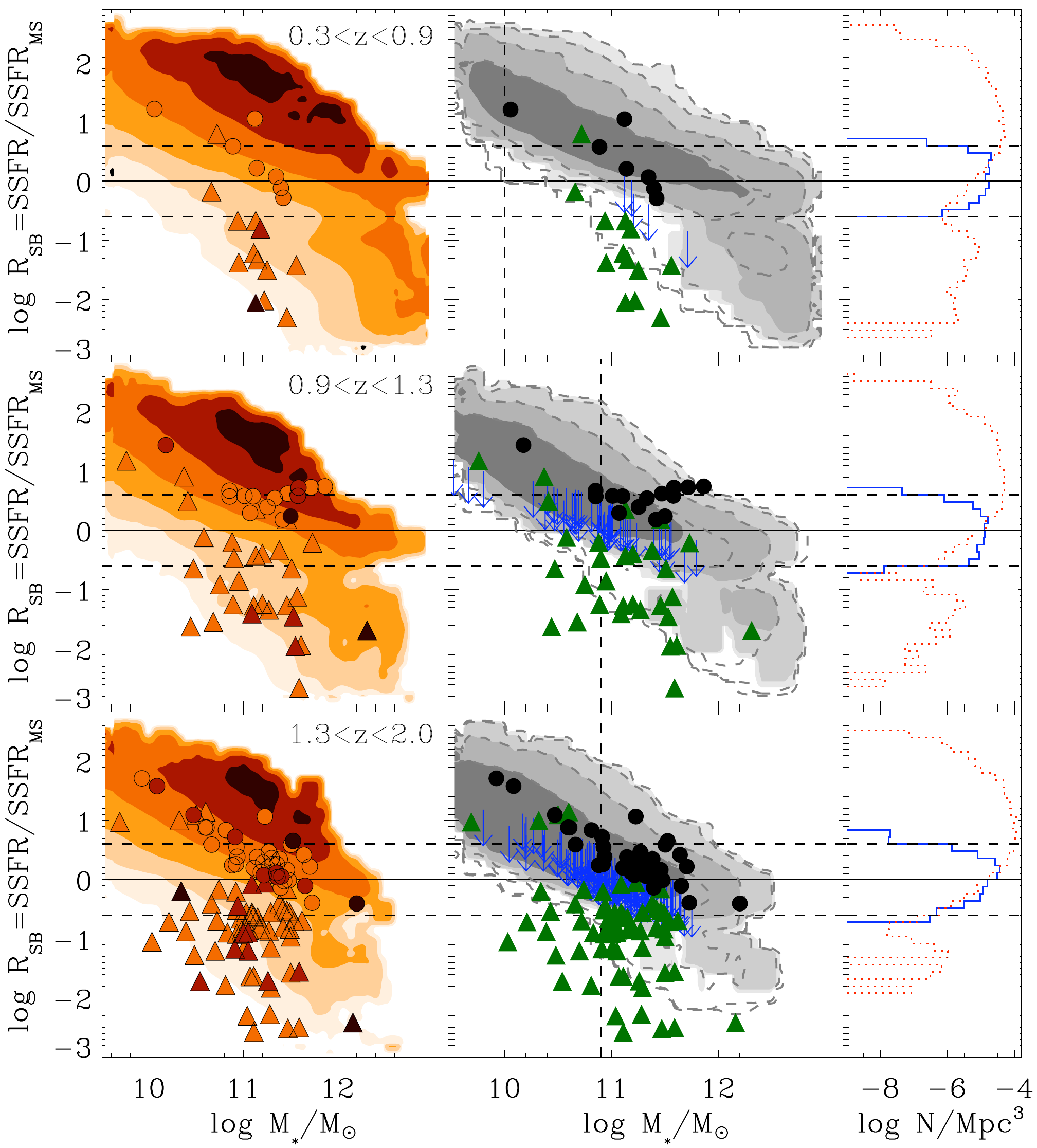}
\caption{Left and central panels: starburstiness $\rm {\rm {R_{SB}}}$  as a function of M$_*$ in three redshift bins. The filled contours in the left panels corresponds to the predicted average values of the AGN X-ray luminosity for bins of different  M$_*$ and $\rm {\rm {R_{SB}}}$. The luminosity values are  equally spaced in logarithmic scale from $\log$ $\rm {L_X}$=42.5 for the lightest filled region to $\log$ $\rm {L_X}$=45  for the darkest. The filled (dotted) contours in the central panel correspond to equally spaced values of the density (per Mpc$^3$) of model AGNs with $\log$ $\rm {L_X}$ $\geq$ 44 erg/s  ($\log$ $\rm {L_X} \ge$ 43.8 erg/s) in logarithmic scale: from 10$^{-9}$  for the lightest filled region to 10$^{-6}$  for the darkest. The data points indicate the XMM-COSMOS AGNs with $\log$ $\rm {L_X} \ge$ 44 erg/s. Circles and arrows indicate AGNs with SFR derived from $\rm{L_{FIR}}$, while triangles indicate AGNs with SFR derived from SED fitting. Circles and triangles are colour coded according to their X-ray luminosity in the left panels.
Solid lines show the position of the galaxy main sequence, while dashed lines denote the limits of the starburst and passive areas, defined as $\rm {\rm {R_{SB}}}>$4 and  $\rm {\rm {R_{SB}}}<$1/4, respectively. Vertical dashed lines indicate the stellar mass  limits  adopted in deriving the fraction of AGN hosted in starburst galaxies in Sect. \ref{fbursty}. Right panels: starburstiness distribution of model AGNs. The solid histograms refer to galaxies dominated by the quiescent mode of star formation (${\rm SFR_{q}}>{\rm SFR_{b}}$), while the dotted histograms refer to galaxies dominated by the burst component of star formation (${\rm SFR_{b}}>{\rm SFR_{q}}$). Solid and dashed lines as in the left panels.}
\label{ssfr_mst}
\end{center} 
\end{figure*}
A step forward in our understanding of  AGN triggering mechanisms can be done by comparing the AGN luminosity with the starburstiness $\rm {\rm {R_{SB}}}$=SSFR/${\rm SSFR_{MS}}$ of the host galaxy \citep{Elbaz11}, where the subscript MS indicates the typical value for main sequence galaxies. The quantity $\rm {\rm {R_{SB}}}$ measures the excess or deficiency in SSFR of a star forming galaxy in terms of its distance from the galaxy main sequence.
 In our previous paper \citep{Lamastra13} we showed that the  correlation between SFR and M$_{*}$ of model  galaxies on the main sequence is determined by  galaxies dominated by the quiescent component of star formation  (${\rm SFR_{q}}>{\rm SFR_{b}}$), while galaxies dominated by the burst component of star formation (${\rm SFR_{b}}>{\rm SFR_{q}}$) have higher values of $\rm {\rm {R_{SB}}}$. Interestingly, we found that the criterion $\rm {\rm {R_{SB}}}>$4, which is commonly assumed to observationally classified starburst galaxies \citep[e.g.][]{Rodighiero11, Sargent12}, works well in  filtering out quiescently star forming galaxies.  Since the more luminous the AGN the larger the burst component of star formation, interaction-driven models for AGNs predict a strong correlation between $\rm {\rm {R_{SB}}}$ and the AGN luminosity. Such a strong luminosity dependence of $\rm {\rm {R_{SB}}}$   is illustrated in figure \ref{ssfr_mst} (left panels), where we show with the contours the average values of the AGN X-ray luminosity as a function of $\rm {\rm {R_{SB}}}$ and M$_*$ in three different redshift bins: 0.3$<z<$0.9,  0.9$<z<$1.3, and 1.3$<z<$2. The figure illustrates that indeed for host galaxies with large values of  $\rm {\rm {R_{SB}}}$  high AGN  luminosities  are expected. An immediate implication of the above is that the fraction of AGN hosts with SSFR above the starburst  threshold ($\rm {\rm {R_{SB}}}>$4) increases with AGN luminosity.

To test this prediction we compare in figure \ref{ssfr_mst}  the luminosity and density distributions in the $\rm {\rm {R_{SB}}}$-M$_*$ diagram of model AGNs (in the central panels the color coding identifies the volume density) with that of X-ray selected AGNs in the XMM-COSMOS field (Section \ref{sample}). 
We restrict the analysis  to the most  luminous sources  with $\rm {L_X} \ge$ 10$^{44}$erg/s. For luminous AGNs the accretion time ($\tau_{AGN}\simeq$10$^7$yrs) is small when compared with the typical lifetime of the SFR indicators based on  UV and IR emission that last $ \sim $ 10$^{8}$ yrs \citep{Neistein13}. The latter time is also longer than the gas depletion time expected from  AGN feedback models based on expanding blast wave. Thus, even if the AGN feedback immediately follows the BH accretion, the signature of star formation in luminous AGNs should be detected even if the galactic gas has been swept out by the AGN feedback.\\ 
Hierarchical clustering models, connecting the properties of galaxies to their merging histories,
reproduce the slope and the scatter of the SSFR-M$_*$ relation; however, they under-predict  its normalization at $ z\lesssim 2$ \citep{Daddi07, Dave08, Fontanot09, Damen09, Santini09, Lin12,Weinmann11,Lamastra13}.  A possible theoretical explanation of this mismatch is that the amount of cold gas in  galaxy disks predicted by these models underestimates the real values.
 In this analysis  we normalize both the model and observed SSFRs to their main sequence values. Both in the model and in the data we obtain $\rm{SSFR_{MS}}$ that depends on redshift and stellar mass. To derive a characteristic $\rm{SSFR_{MS}}$ at a given redshift and stellar mass,  for model galaxies we  separately fit\footnote{We use the spline IDL routine}  the peaks of the SFR distributions  as a function of the stellar mass  with the relation $\log \rm{SFR}=a\log\rm{M_*}+b$ in each individual redshift bins. We found values for (a,b) equal to (0.91,-9.02), (0.96,-9.31), and (0.93-8.76) in the three redshift bins from $z=$0.3 to $z=$2, respectively. For the observational data we use the  best-fit of the galaxy main sequences obtained by \cite{Santini09} in similar redshift intervals. These relations have slopes flatter than those obtained for model galaxies ranging from 0.65 to 0.85.\\
 \begin{figure}[h!]
\begin{center}
\includegraphics[width=5 cm]{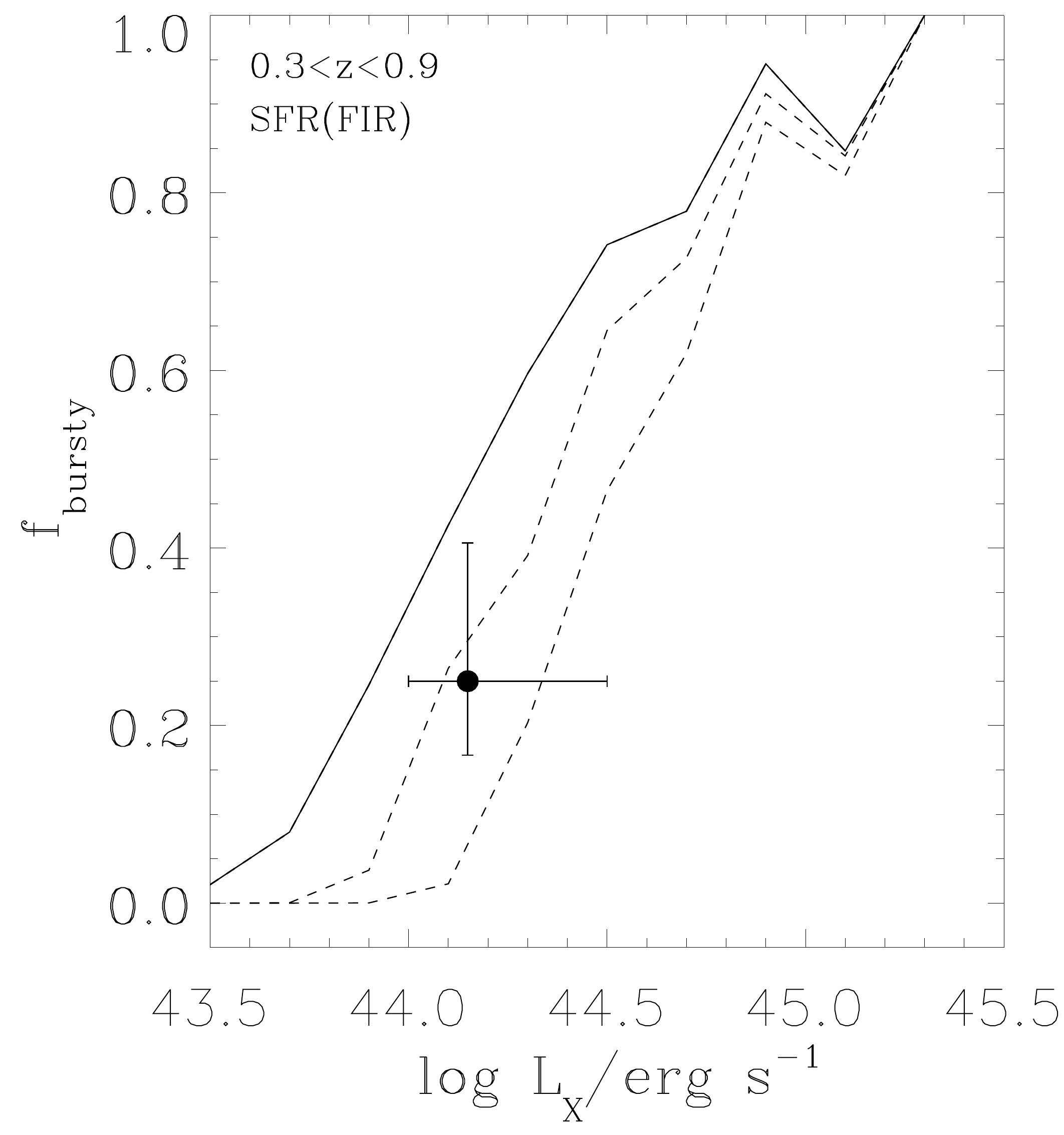}
\includegraphics[width=5 cm]{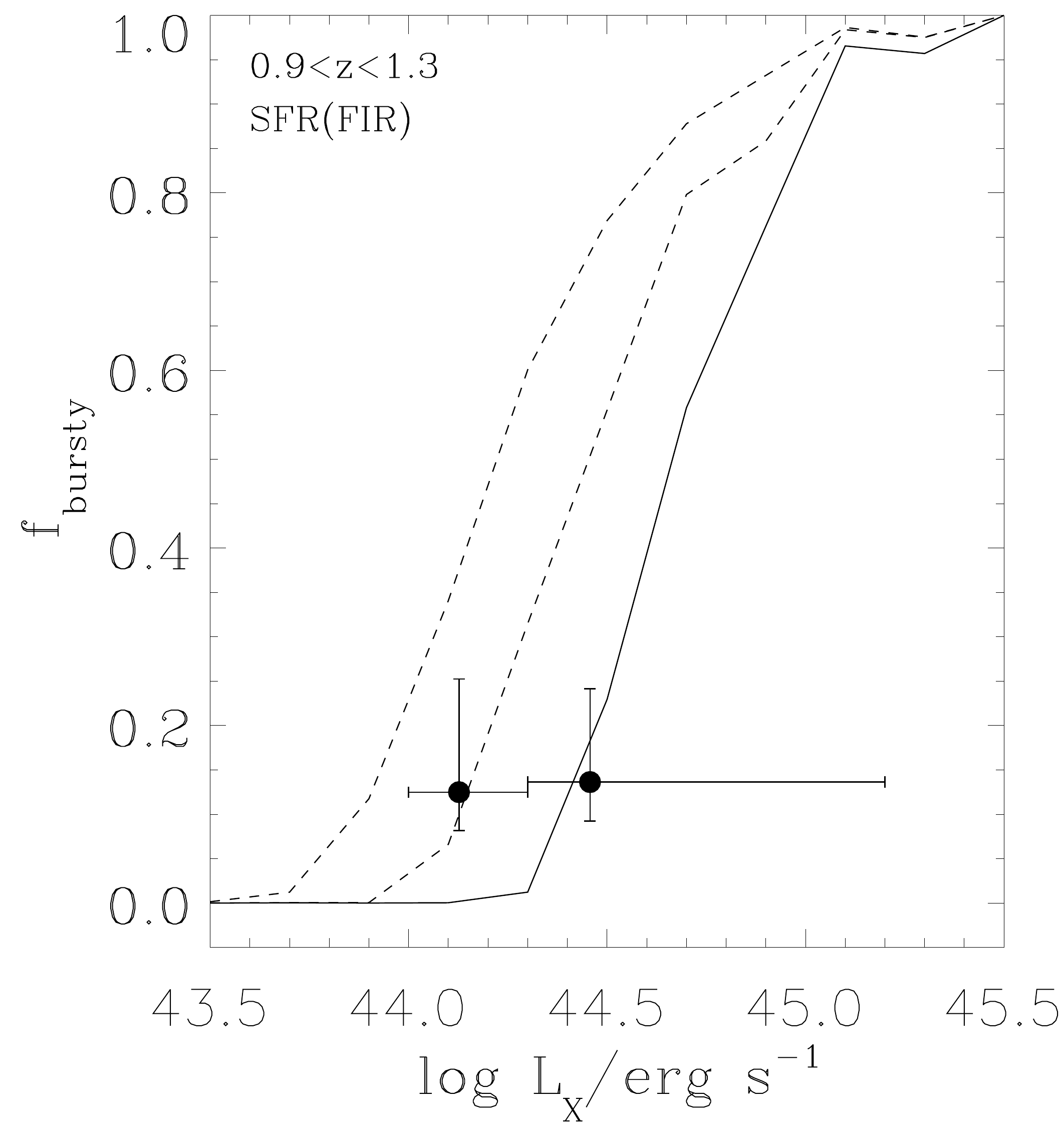}
\includegraphics[width=5 cm]{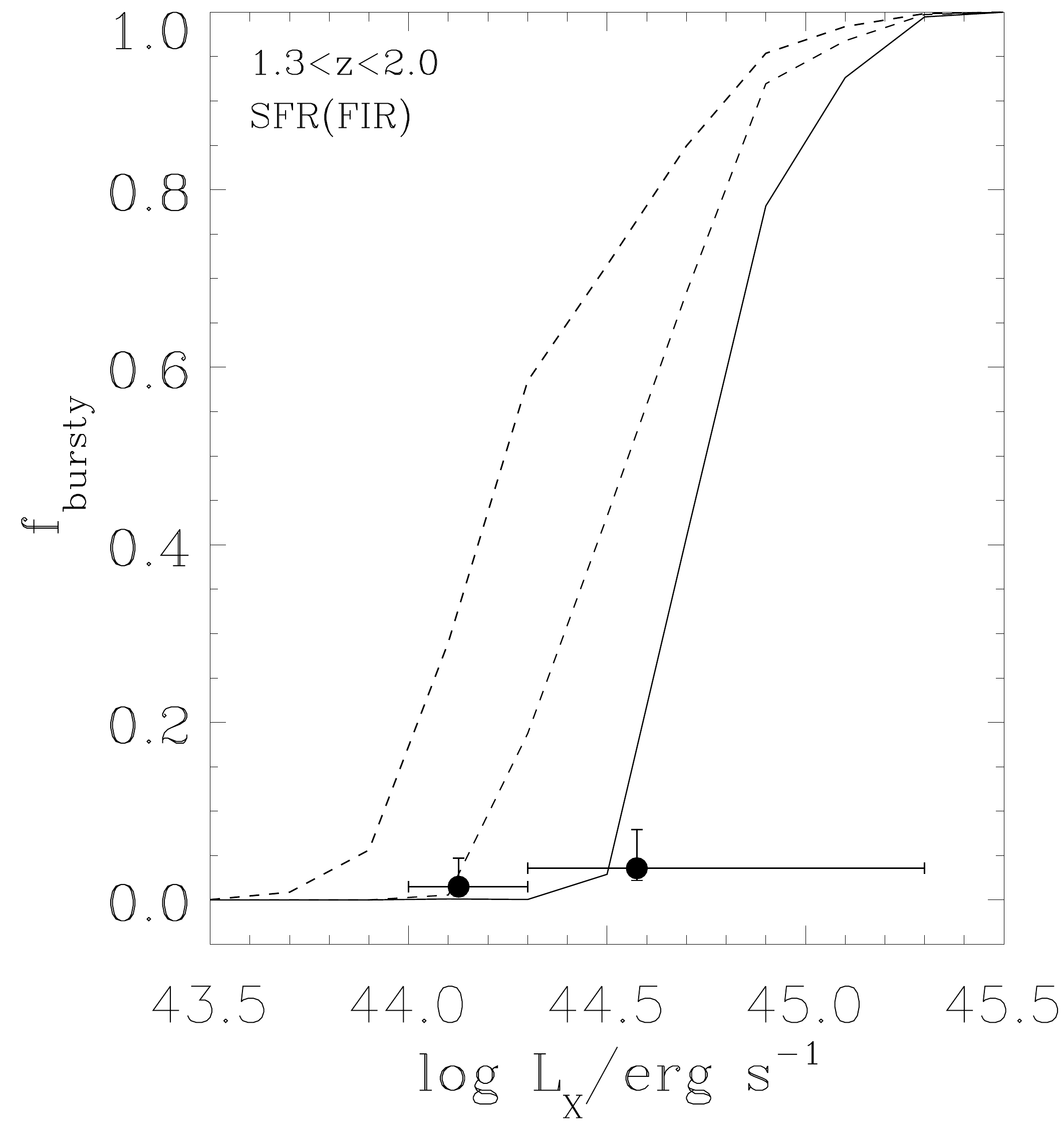}
\caption{$\rm {\rm {f_{bursty}}}$ versus $\rm {L_X}$ in three redshift bins. The lines show the model predictions obtained by selecting AGNs with M$_* \geq 10^{10} \rm{M_{\odot}}$, M$_* \geq 10^{10.5} \rm{M_{\odot}}$, and M$_* \geq 10^{10.9} \rm{M_{\odot}}$ from left to right. Solid lines indicate the stellar mass limits used to derive the observational fractions. The data points are derived using the FIR-based SFRs and M$_*$ derived  by \cite{Santini12}. The plotted value of $\rm {L_X}$ is the median value for sources in each luminosity bin. Vertical error bars indicate the 1$\sigma$ binomial uncertainties. Horizontal error bars indicate the luminosity bin sizes which are optimized to have roughly the same number of sources in each bin. }
\label{fbursty_Lx}
\end{center} 
\end{figure}
As it can be seen in figure \ref{ssfr_mst} (central panels) the model predicts that  low mass galaxies (M$_* \lesssim$ 10$^{10} {\rm M_{\odot}}$) hosting bright AGNs ($\rm {L_X} \ge$ 10$^{44}$erg/s) are predicted to populate the starburst region ($\rm {\rm {R_{SB}}}>$4), while higher stellar mass hosts  mainly populate the main sequence (1/4$<\rm {\rm {R_{SB}}}<$4) and the passive areas ($\rm {\rm {R_{SB}}}<$1/4). The effectiveness of the $\rm {\rm {R_{SB}}}>$4 criterion in  filtering out quiescently star forming galaxies can be inferred  from  the right panels of figure \ref{ssfr_mst},  which show separately the starburstiness distributions  of AGN  hosts dominated by the quiescent component of star formation, and of AGN hosts dominated by the burst component of star formation. Indeed, all the galaxies in the starbust  region are  dominated by the burst component of star formation. The main sequence region is nearly equally populated by AGN hosts with ${\rm SFR_{b}}>{\rm SFR_{q}}$ and  with ${\rm SFR_{q}}>{\rm SFR_{b}}$, while the passive area is populated by massive galaxies dominated by the burst component of star formation. Such massive hosts formed from biased, high density regions of the primordial density field where the frequent high-$z$ interactions rapidly convert the cold gas reservoir into stars at early cosmic epochs, leaving only a residual fraction of cooled gas available for the star formation   at $z \lesssim$2. These massive galaxies represent only the 1\% of the $\rm {L_X} \ge$ 10$^{44}$erg/s  AGN hosts at  $z>$0.9, at lower redshift this fraction increases up to 10\%.

We also show in figure \ref{ssfr_mst}  the model predictions obtained by selecting from the model AGNs with $\rm {L_X} \ge$ 10$^{43.8}$erg/s (dotted contours in the central panels). Indeed, the uncertainty related to the estimate of the intrinsic AGN X-ray luminosity due to the  uncertainty in the bolometric correction\footnote{for model AGNs we use the luminosity-dependent bolometric correction factor given in \cite{Marconi04} to derive  X-ray luminosities in the 2-10 keV band from bolometric luminosities.}  and to uncertainty in  AGN absorbing column density measurements (especially from hardness ratio) can affect the comparison between the model and the data. We note that a larger discrepancy in the luminosity distribution in the $\rm {\rm {R_{SB}}}$-M$_{*}$ diagram  between the data and the model  is observed for the obscured AGN sample (see left panels of figure \ref{ssfr_mst}).\\ 
By comparing the density distribution of model AGNs with that of the XMM-COSMOS AGNs with $Herschel$ observations (figure \ref{ssfr_mst}, central panels), we find that all FIR-detected AGNs  lie in the predicted confidence region represented by the contour plots. However, this  comparison is hampered by the $Herschel$ detection limit in the COSMOS field (corresponding to SFR limits of $\sim$6 $\rm {M_{\odot}}$/yr  at $z=0.3$ and  $\sim$400$\rm {M_{\odot}}$/yr  at  $z=2$) that allows to estimate only SFR upper limits for a large number of the sources.  For about 30\%  (59/196) of the $Herschel$-undetected sources SFR estimates from SED fitting in the optical/IR band are available. Only for a very small fraction of these sources (2/59)   160 $\mu$m upper limits are not consistent with the SED-based estimates.
By comparing the model predictions with the obscured AGN sample with SED-based SFRs we do not find an exact overlap with the confidence regions predicted by the model in the passive areas. Indeed, the model predicts that the  hosts of luminous AGNs with low star formation are more massive than observed, especially in the  lowest redshift bin. This may be related to an incompleteness in our treatment of the AGN feedback which is a mechanism of suppression of the cooling in massive halos. Such feedback must be still at work at low redshift to continuously suppress star formation in massive halos at $ z<$1.  In fact, such long-standing problem of the hierarchical scenario of galaxy formation lead to the over-prediction of high mass galaxies in the local Universe as shown by comparing ours and other SAMs with the observed stellar mass function (e.g. \citealt{Menci05,Fontanot09, Guo11}, but see \citealt{Bernardi13} and \citealt{Mitchell13}).

However,  the mismatch between the data and the model at low SSFR values could be at least partially explained by the systematics affecting the SFR indicators. \cite{Bongiorno12} compared the SFRs computed using FIR data and those computed with the SED fitting  for the AGNs in the COSMOS field finding that the SFR computed from the SED are systematically lower than the one derived from the FIR \citep[see fig. B1 of][]{Bongiorno12}. They concluded that this discrepancy is the result of a combination of two effects: (i) the tendency for the SED fit to overestimate the AGN emission component, (ii) the FIR overestimation of the SFR, especially at high AGN luminosities and low SFR, where the AGN contamination in the IR band is not negligible.\\

In order to study the dependence of the starburstiness  of the host galaxy on the AGN luminosity we  estimate the fraction $\rm {\rm {f_{bursty}}}=N_{AGN}^{SB}/N_{AGN}$ of AGN host galaxies with $\rm {\rm {R_{SB}}}>$4  relative to the total number of AGN hosts as a function of $\rm {L_X}$. The predictions from the model for different host galaxy stellar masses are shown as lines in  figure \ref{fbursty_Lx}. As expected,  a strong correlation is predicted by the model for high AGN luminosities ($\rm {L_X} \ge$ 10$^{44}$erg/s).  The fraction $\rm {\rm {f_{bursty}}}$ is predicted to increases rapidly  from $ \lesssim $0.2 at $\rm {L_X} \lesssim$ 10$^{44}$erg/s to $ \gtrsim$0.9 at $\rm {L_X} \gtrsim$ 10$^{45}$erg/s. \\
In this figure we compare the model predictions with the observational results obtained from the FIR-based SFRs and the stellar masses derived by \cite{Santini12}. The estimate of $\rm {\rm {f_{bursty}}}$ from FIR data is prevented by the large number of $Herschel$ undetected sources. However, we note that,  in each redshift bin above a given stellar mass all AGNs in the starburst region are detected by $Herschel$ (see  figure \ref{ssfr_mst}). Thus, above these stellar masses we can properly estimate $\rm{N_{AGN}^{SB}}$ and hence $\rm {\rm {f_{bursty}}}$. These stellar mass limits depends primarly on the instrumental sensitivity which corresponds to larger SFR detection limits at higher redshifts, and on the evolution of the galaxy main sequence. For the evolution adopted in this paper  they correspond  to M$_* \geq 10^{10} \rm{M_{\odot}}$ at 0.3$<z<$0.9, and to M$_* \geq 10^{10.9} \rm{M_{\odot}}$ in the higher redshift bins.
The fractions $\rm {\rm {f_{bursty}}}$  are computed in $\rm {L_X}$ intervals optimized to have roughly the same
number of sources in each bin. The results are shown in figure \ref{fbursty_Lx}, where, in
each luminosity interval, the plotted value of $\rm {L_X}$ is the median value in
the bin and 1$\sigma$ uncertainties are derived through binomial statistics. For $\rm {L_X} \lesssim$10$^{44.5}$ erg/s the data are 
 in reasonable good agreement with the model predictions, except for AGNs with $\rm {L_X} \simeq$10$^{44.1}$ erg/s at 0.9$<z<$1.3 for which the model predicts a lower value of $\rm {\rm {f_{bursty}}}$ than that observed. The measurements of the SFR and stellar mass of AGN hosts in  larger area surveys  are necessary  to probe with the required statistics the high-luminosity range ($\rm {L_X} \ge$ 10$^{44.5}$erg/s) where the strongest dependence of $\rm {\rm {f_{bursty}}}$  on $\rm {L_X}$ is expected.\\
We also tested the robustness of our results by adopting different galaxy main sequences for the observational data. By adopting the galaxy main sequence  from \cite{Whitaker12}, and a $\rm {SFR_{MS}}$ that varies with stellar mass with a slope of 0.8 \citep[e.g.][]{Rodighiero11} and evolves with time as (1+$z$)$^{2.95}$ \citep[e.g.][]{Pannella09,Elbaz11,Magdis12},   we found consistent results within the errors, except 
at 0.9$<z<$1.3 once the latter parametrization is assumed. For this redshift range  we found  $\rm {\rm {f_{bursty}}}$=0 in each luminosity bin. This can be explained by the steeper slope of the galaxy main sequence that corresponds to  larger $\rm {SSFR_{MS}}$ for galaxies with M$_* \gtrsim 10^{10} \rm{M_{\odot}}$.
  \\

\subsubsection{The role of obscuration}
\begin{figure}[h!]
\begin{center}
\includegraphics[width=5 cm]{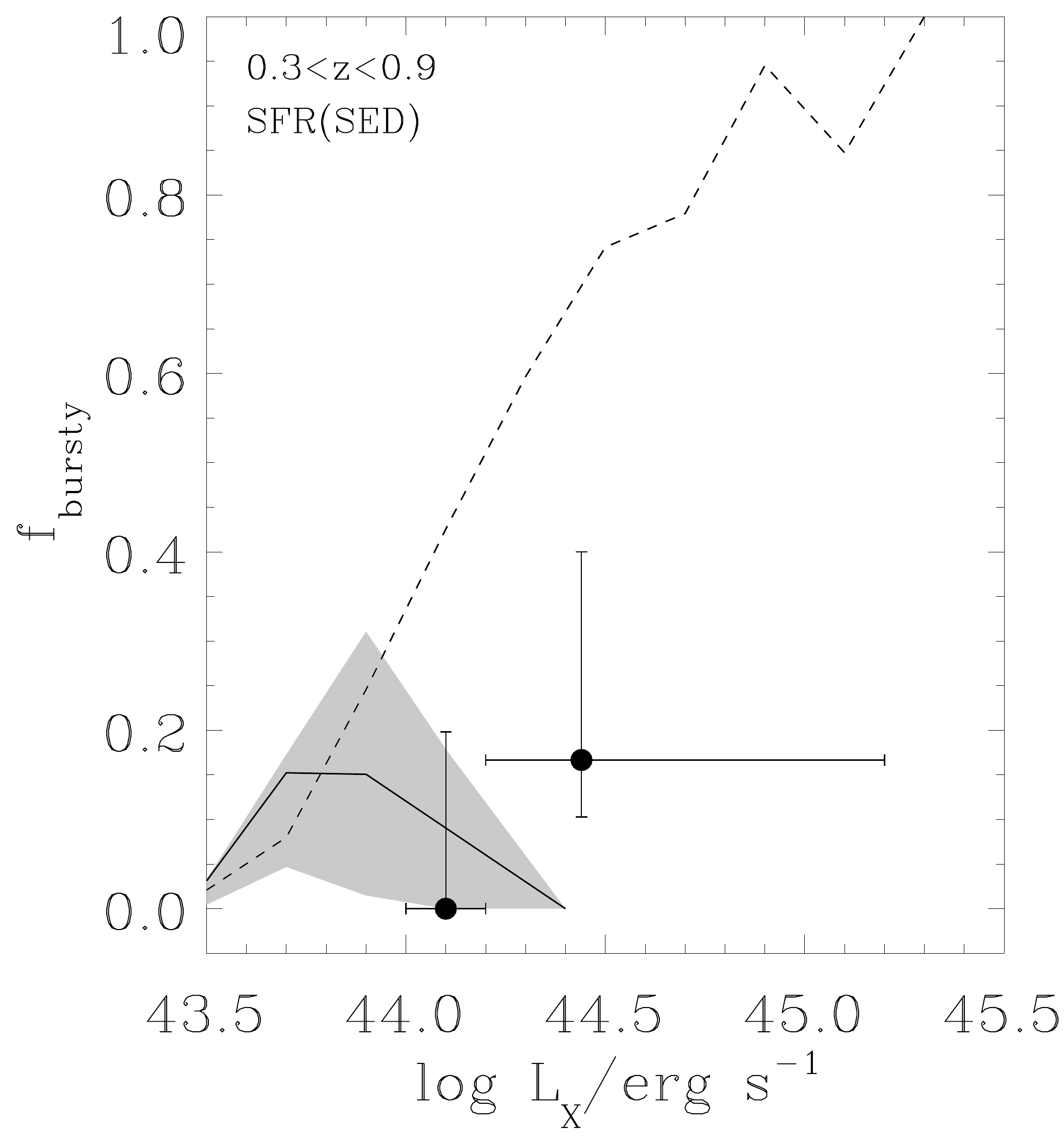}
\includegraphics[width=5 cm]{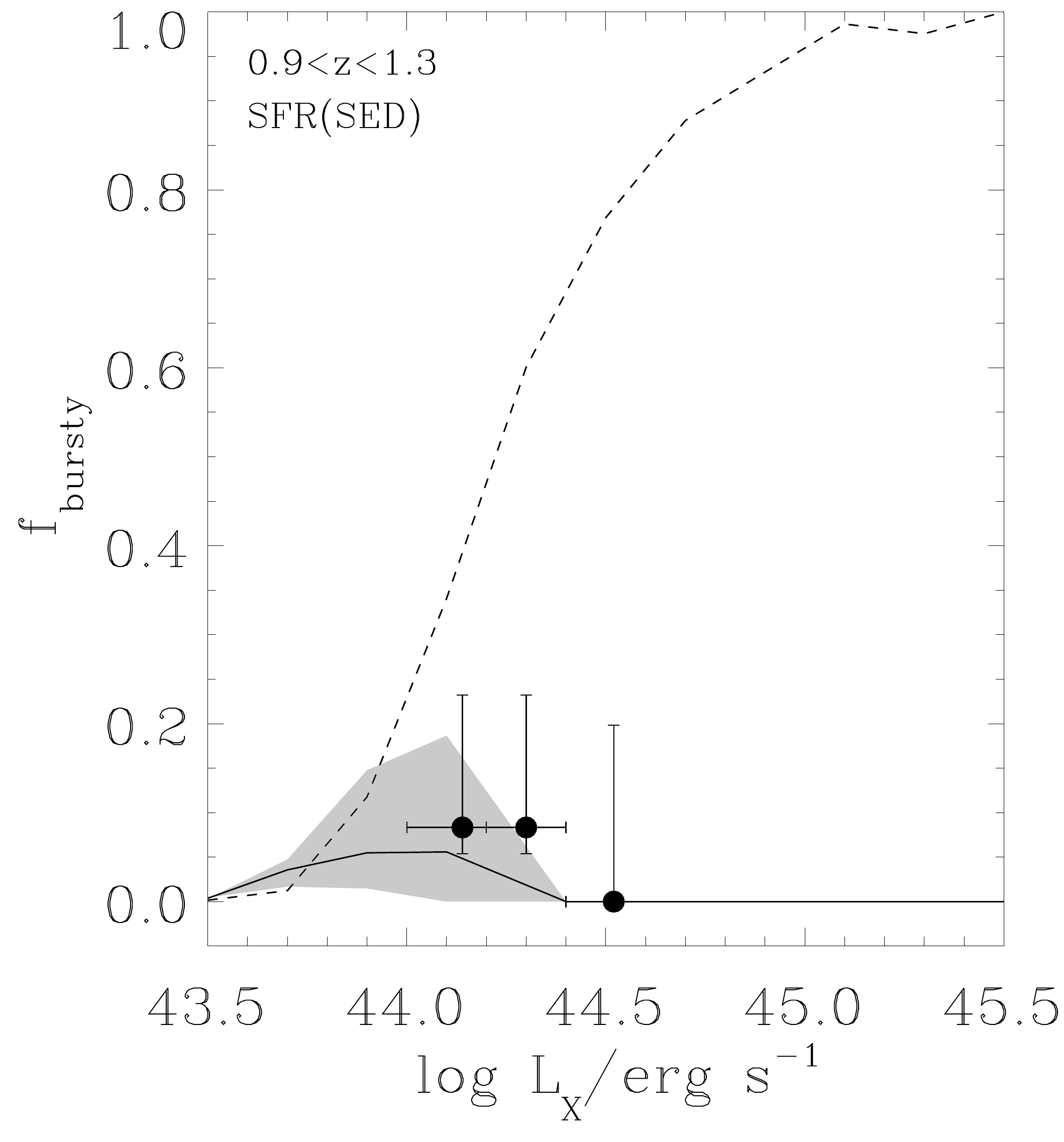}
\includegraphics[width=5 cm]{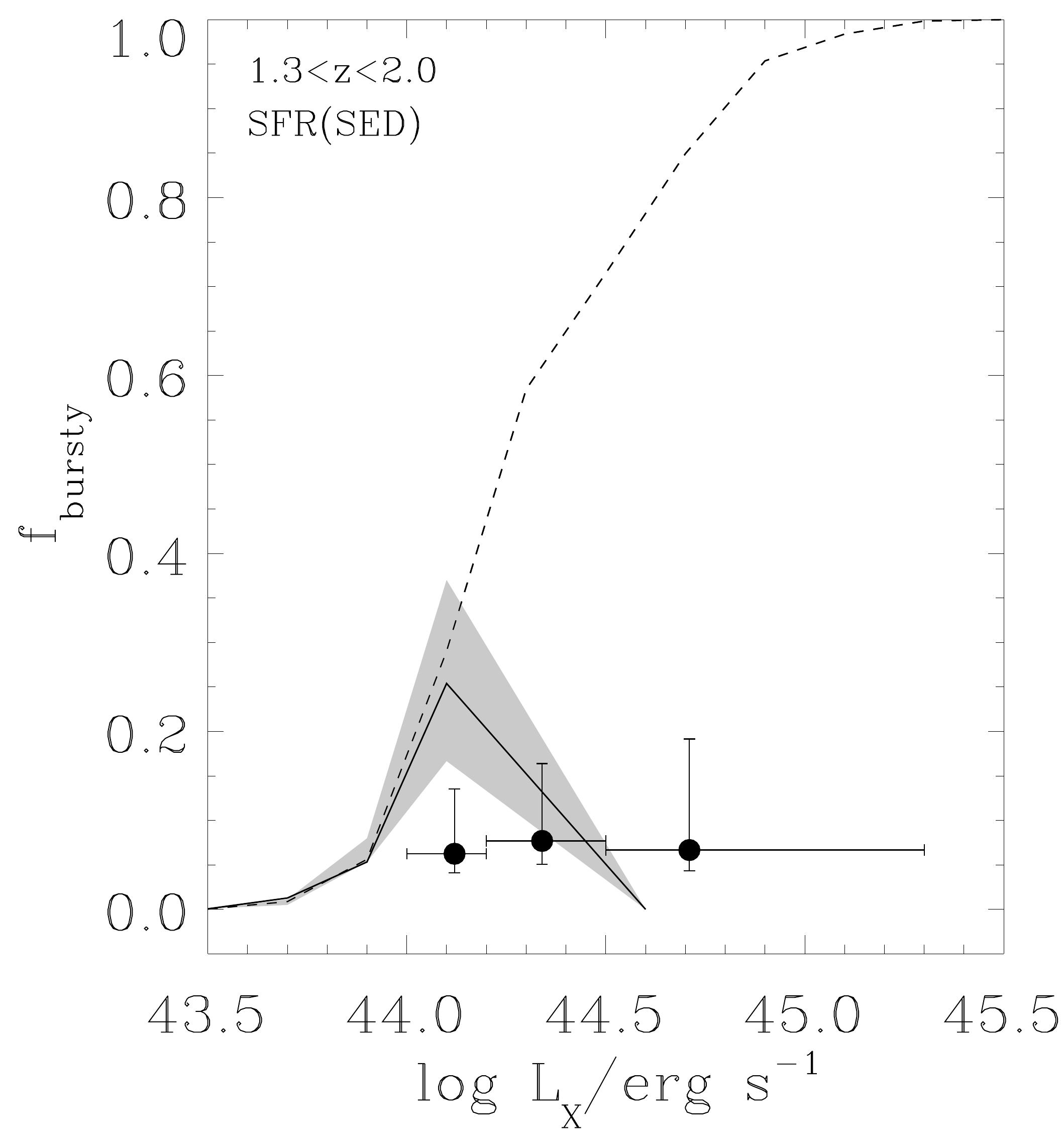}
\caption{$\rm {\rm {f_{bursty}}}$ versus $\rm {L_X}$ in three redshift bins. The solid lines show the model predictions obtained by selecting obscured AGNs with $\log$ $\rm {N_H} >$22 cm$^{-2}$ and  M$_* \geq 10^{10} \rm {M_{\odot}}$. The upper and the lower envelopes of the shaded regions show $\rm {\rm {f_{bursty}}}$  corresponding to the selection
$\log$ $\rm{N_{H}} \geq$21.8 cm$^{-2}$ and $\log$ $\rm{N_{H}} \geq$22.2 cm$^{-2}$, respectively. The dashed lines show the predictions obtained by selecting obscured and unobscured  AGNs with  M$_* \geq 10^{10} \rm {M_{\odot}}$.
The data points  are derived using the SED-based SFRs and M$_*$ as computed  by \cite{Bongiorno12}. The plotted value of $\rm {L_X}$ is the median value for sources in each luminosity bin. Vertical error bars indicate the 1$\sigma$ binomial uncertainties. Horizontal error bars indicate the luminosity bin sizes which are optimized to have roughly the same number of sources in each bin.}
\label{fbursty_Lx_type2}
\end{center} 
\end{figure}

In the previous section we showed that the $Herschel$ sensitivity in the COSMOS field allows us to derive  $\rm {\rm {f_{bursty}}}$ only for host galaxies with large stellar masses.
In order to extend this analysis to lower stellar mass hosts, in this section we derive $\rm {\rm {f_{bursty}}}$ using the SFRs derived  from the SED fitting technique. This allow us  to derive an independent estimate of $\rm {\rm {f_{bursty}}}$. However, the SED-based SFRs restricts the analysis only to  obscured AGNs (see Section \ref{sample}). 
Using the SFRs and the stellar masses derived by \cite{Bongiorno12} we estimate $\rm {\rm {f_{bursty}}}$ as a function of $\rm {L_X}$  (in the usual three redshift bins) as computed in the previous section.
The results are shown with the data points in figure \ref{fbursty_Lx_type2}. These fractions remain almost unchanged if the different parametrizations of the galaxy main sequences described in the previous section are adopted.\\
As a comparison, we show the model predictions obtained by selecting only obscured AGNs. We use  the canonical absorbing column density  $\log$ $\rm{N_{H}} \geq$22 cm$^{-2}$ to select from the model obscured AGNs. However the exact values of $\rm {\rm {f_{bursty}}}$
predicted by the model depend on the $\rm{N_{H}}$ threshold adopted  as it is indicated by  the upper and the lower envelopes of the shaded regions which show the fractions obtained by selecting AGNs with 
$\log$ $\rm{N_{H}} \geq$21.8 cm$^{-2}$ and $\log$ $\rm{N_{H}} \geq$22.2 cm$^{-2}$, respectively.\\
We compare the model predictions with the observations, with the caveat that in the model the obscuration is associated only to cold gas in the galaxy disk, while the observational classification is based on both nuclear and galactic obscuration \citep[see][for the details about the AGN classification]{Bongiorno12}. The model predictions are in reasonable good agreement with the observational data in the luminosity ranges where obscured AGNs are predicted. However, the observations indicate the presence of obscured AGNs also at higher luminosities than those predicted. 

A striking feature of the model is that at high X-ray luminosities the predicted  dependence of $\rm {\rm {f_{bursty}}}$ on  $\rm {L_X}$  is different  for the obscured and unobscured AGN populations.
Indeed, for the latter, $\rm {\rm {f_{bursty}}}$  is a monotonically  increasing function of  $\rm {L_X}$, while for obscured AGNs $\rm {\rm {f_{bursty}}}$ initially increases with $\rm {L_X}$ (similarly to the unobscured population) and then decreases. 
This is due to the fact that obscured AGNs correspond to early stages of feedback action; in particular for a given orientation of the line of sight, the observed column density depends on the time elapsed since the start of the blast wave expansion. The faster expansion characterizing the blast wave of luminous AGNs thus corresponds to a larger probability that we will observe them as unobscured AGNs. Thus, the predicted fraction of obscured AGN decreases with increasing AGN luminosity \citep{Menci08}. Moreover, the probability of finding a luminous AGN in a gas rich galaxy is low for galaxies with high values of SFR owing to the  energy released into the interstellar medium by SNae  feedback.\\

\subsection{The starburstiness of  high-$z$ QSOs}

\begin{figure*}[t]
\begin{center}
\includegraphics[width=7 cm]{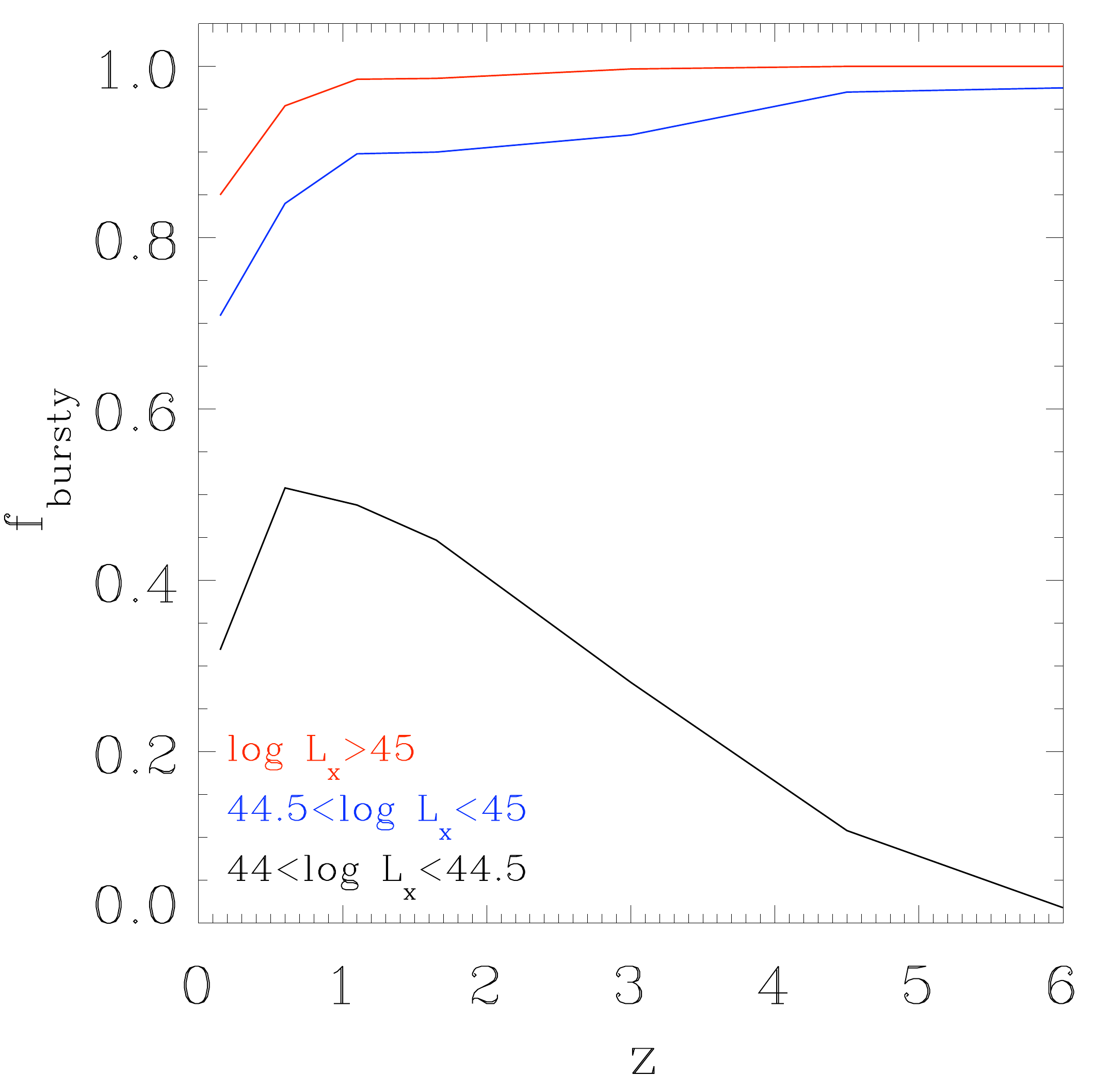}
\includegraphics[width=7 cm]{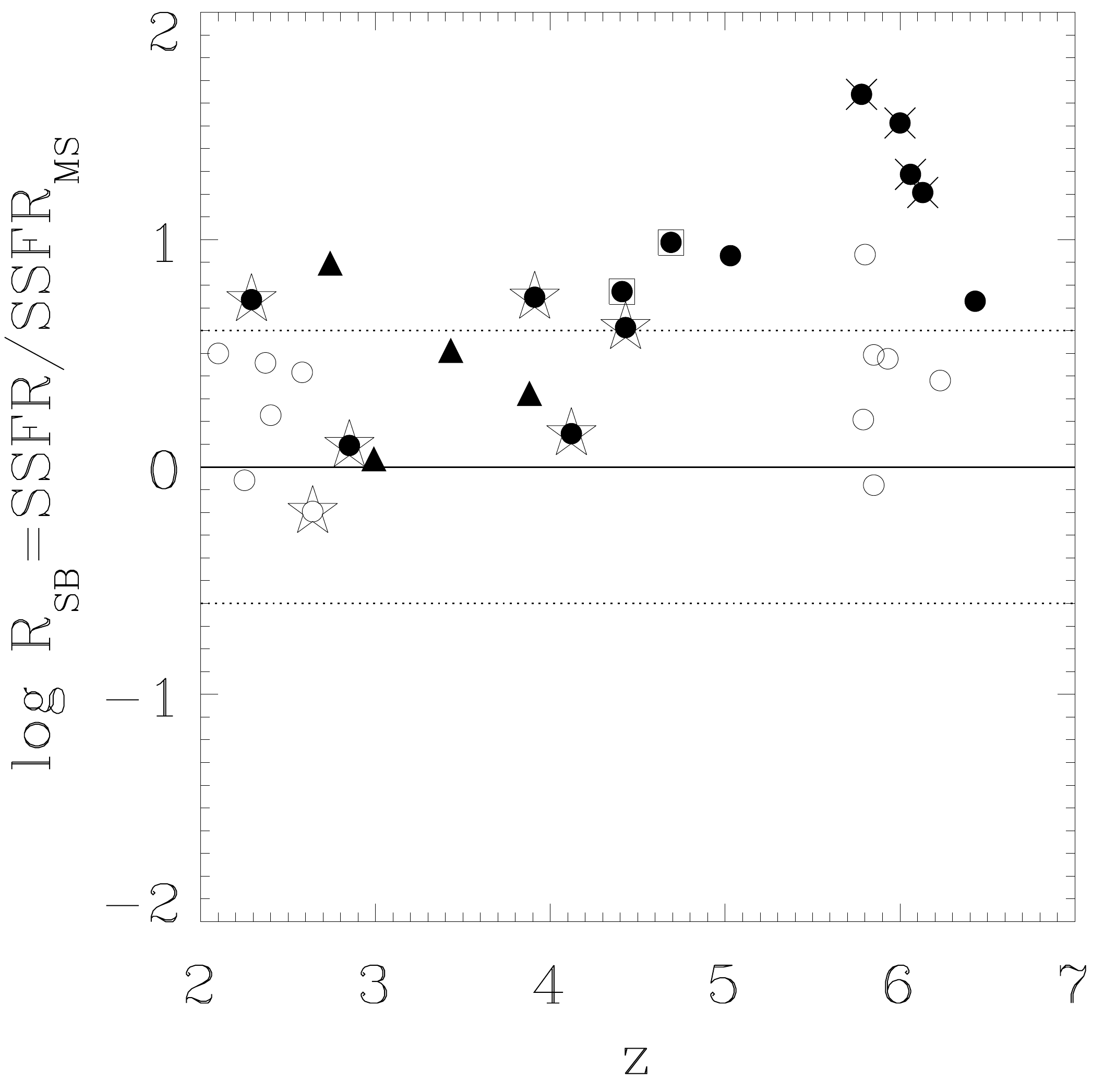}
\caption{Left: $\rm {\rm {f_{bursty}}}$  as a function of $z$ for three different AGN luminosity bins: 44$<\rm {\log L_{X}}<$44.5, 44.5$<\rm {\log L_{X}}<$45, and $\rm {\log L_{X}}>$45 from bottom to top. Right:
The starburstiness  $\rm {\rm {R_{SB}}}$ as a function of $z$ of QSOs with  $\rm {L_{bol}} >$10$^{46}$ erg/s. Triangles: stellar masses from SED fitting \citep{Polletta08,Lacy11}; circles: stellar masses from dynamical masses within the CO emitting region  \citep{Wang10,Solomon05,Coppin08,Shields06,Maiolino07,Gallerani12b} and the [CII] emitting region (crossed circles,\cite{Wang13}. For the sources with [CII] measurements $\rm{M_{dyn}}$ are calculated assuming the disk inclination angle estimated from the [CII] minor and major axis ratio \citep[see][]{Wang13}. Open circles indicate dynamical masses derived assuming a disk radius of 2-2.5 Kpc , while filled circles indicate dynamical masses obtained from spatially resolved measurements of the molecular gas emitting region. Squared circles denote sources in which $R$ is measured as half the component separation in merger model \citep[see][]{Shields06}. Starred circles denote  gravitationally lensed QSOs for which the CO and FIR  luminosities have been corrected for magnification \citep{Riechers11}.}
\label{fbursty_z}
\end{center} 
\end{figure*}

The results shown in the previous sections indicate that the measurements of SFR and  M$_*$ of luminous AGNs are fundamental to constrain AGN triggering mechanisms.
Indeed, interaction-driven models for AGNs predict that  a large fraction ($ \gtrsim $0.8) of galaxies hosting  high-luminosity  AGNs ($\rm {\log L_{X}} \gtrsim$44.5) have SSFRs large enough to be selected as starburst ($\rm {\rm {R_{SB}}}>$4).
This is valid at all epochs, as it is shown in figure \ref{fbursty_z} (left panel) where the predicted $\rm {\rm {f_{bursty}}}$ is given as a function of $z$ for three different  AGN luminosity bins.

At $ z\lesssim $0.5 the evolution of $\rm {\rm {f_{bursty}}}$   is similar for AGNs with different luminosity. The increase of $\rm {\rm {f_{bursty}}}$  with $z$ is due to the decrease of the fraction of  massive galaxies (M$_* \geq 10^{12} \rm {M_{\odot}}$)  hosting  AGNs  with these luminosities  (see fig. \ref{ssfr_mst}). This similar evolution  implies that the slope of the $\rm {\rm {f_{bursty}}}$-$\rm {L_X}$ relation remains almost unchanged  up to $ z\simeq $0.5.

At higher redshifts the $\rm {\rm {f_{bursty}}}$-$\rm {L_X}$ relation steepens. In fact,  the fraction of starbursting systems hosting  AGN with  $\rm {\log L_{X}}>$44.5 remains nearly constant with $z$, while  the fraction of starburst galaxies hosting lower luminosity  AGNs  decreases with redshift. The latter trend is determined by the increase of the normalization of the galaxy main sequence with redshift. The predicted evolution of the galaxy main sequence, which  determines the position of  the peak at $z\simeq $0.5 of the  $\rm {\rm {f_{bursty}}}$ evolution for  low-luminosity AGNs, is driven by the larger amount of cold gas  available for star formation at earlier epochs and by the shorter star formation time scale $\tau_q \propto t_d$ of high-$z$ galaxies.
Despite the increase of the main sequence's normalization with redshift, the model predicts that galaxies hosting high-luminosity AGNs lie well above the galaxy main sequence at all redshifts.
To test this prediction we estimate the starburstiness of the high-$z$ QSOs with $\rm {L_{bol}} \geq $10$^{46}$erg/s ($\rm {\log L_{X}}\gtrsim$44.5 erg/s) belonging to the heterogeneous sample described in Section \ref{sample}. 
To estimate $\rm {\rm {R_{SB}}}$ in these objects we conservatively adopted SFRs derived by assuming that the 
IR luminosity is dominated by the FIR emission in these objects and considering the AGN contribution in the FIR  (Section 3). 
To estimate the stellar masses of the CO-detected QSOs from the difference between dynamical and  gas masses,  for the latter we neglect the contribution of the atomic gas and we assume 
$\alpha_{C0}$=0.8 (K km s${-1}$ pc$^2$)$^{-1}$. This value  is lower than the one used for disk galaxies \citep[$\alpha_{CO}\gtrsim$4 (K km s${-1}$ pc$^2$)$^{-1}$, e.g.][]{Daddi10,Genzel10,Magdis11,Magnelli12}.  Although the exact value of  $\alpha_{CO}$ for starburst and disk galaxies is a debated topic \citep[see][for a review]{Bolatto13}, the value of  $\alpha_{CO}$ assumed in this work  corresponds  to lower gas masses for a given CO line luminosity for QSO hosts than for disk galaxies.  Thus our estimates of the stellar masses represent upper  limits.
We use the  galaxy main sequences obtained by \cite{Santini09}, \cite{Daddi09}, and \cite{Stark09}  for the redshift intervals 2$ <z< $2.5, 2.5$ <z< $4, and $ z> $  4, respectively. The resulting $\rm {\rm {R_{SB}}}$  is shown  as function of $z$ in figure \ref{fbursty_z}  (right panel).
We find that  $ \sim $90\% (26/29) of these QSOs  lies above the galaxy main sequence and  that $ \sim $45\% (13/29) has $\rm {\rm {R_{SB}}} \geq$4.
Although  the latter fraction is lower than that predicted by the model ($ \gtrsim $0.8, see left panel of \ref{fbursty_z}), it represents a lower limit owing to the conservative SFR and M$_*$ estimates that we have adopted. 
This analysis suggests that on average the host of luminous QSO are more ``starbursty" than normal star forming galaxies, bearing in mind the large uncertainties in the estimation of the stellar masses in these objects. A similar trend is also found  for radio AGNs at $z<$2 \citep[][]{Karouzos13}.

\section{Conclusions}\label{conclusions}
We have investigated the star formation properties of the hosts of luminous ($\rm {L_X} \geqslant$10$^{44}$ erg/s) AGNs predicted under the assumption that starburst events and AGN activity are triggered by galaxy encounters during their merging histories. The latter are described through Monte Carlo realizations and are connected to  star formation and BH accretion using an SAM of galaxy formation
in a cosmological framework. We compared the model predictions with new measurements of  the fraction of  AGNs hosted in starburst galaxies as a function of the AGN luminosity  in the redshift range  0.3$< z< $6.5 to constrain  AGN triggering mechanisms.
The main results of this paper follow.
\begin{itemize}

\item [$\bullet$] Pinning down the AGN triggering mechanism from the relation between the SFR of the host galaxy and the AGN luminosity is a difficult task. On the observational side, the observed relation is still uncertain due to observational bias and instrumental limitation in sensitivity. On the theoretical side, even a model based on galaxy interactions as triggers of AGN and starburst activities, predicts a  large scatter ($\simeq$3 orders of magnitude) at low AGN luminosities  owing to the large contribution of the quiescent component of star formation to the total SFR of the galaxy in these sources.

\item [$\bullet$] The relation between  the  AGN luminosity and the fraction $\rm {\rm {f_{bursty}}}$ of AGNs hosted in starburst galaxies is a powerful tool to constrain AGN triggering mechanisms since
the starburstiness $\rm {\rm {R_{SB}}}$=SSFR/${\rm SSFR_{MS}}$ of the host galaxy is an effective  diagnostic to separate  the quiescent and starburst  modes of star formation \citep{Lamastra13}.  By adopting a starburst threshold of $\rm {\rm {R_{SB}}}>$4 \citep{Rodighiero11,Sargent12} we find that the predicted fraction $\rm {\rm {f_{bursty}}}$  increases with AGN X-ray luminosity  from $ \lesssim $0.2 at $\rm {L_X} \lesssim$ 10$^{44}$erg/s to $ \gtrsim$0.9 at $\rm {L_X} \gtrsim$ 10$^{45}$erg/s over a wide redshift interval from $ z\simeq $0 to $ z\simeq $6.

\item [$\bullet$] Interaction-driven models including AGN feedback related to the luminous AGN phase  predict  that at low X-ray luminosities ($\rm {L_X} \lesssim$ 10$^{44}$erg/s) $\rm {\rm {f_{bursty}}}$ increases with  $\rm {L_X}$ similarly for unobscured and obscured AGNs, while at higher luminosities $\rm {\rm {f_{bursty}}}$ steeply increases and decreases for the unobscured and obscured populations, respectively.

\item [$\bullet$] The sharp, steep relation between $\rm {\rm {f_{bursty}}}$ and the AGN luminosity predicted by interaction-driven models implies that a large fraction ($\simeq$80\%)  of luminous AGNs ($\rm {L_X} \ge$ 10$^{44.5}$erg/s) are in starburst galaxies. At present, observations indicate that at least $\simeq$50\%  of the QSO hosts at 2$< z< $6.5  are starburst galaxies. Future systematic studies of the stellar properties of high luminosity AGNs are therefore necessary in order to make a step forward in our understanding of  AGN triggering mechanisms.


\end{itemize}

\section*{Acknowledgements} The authors thank Benjamin Magnelli for  kindly providing the SFR of HS1700+6416, and the referee for helpful comments. This work was supported by ASI/INAF contracts I/024/05/0 and I/009/10/0 and PRIN INAF 2011, 2013.

\bibliographystyle{aa}
\bibliography{biblio.bib}

\end{document}